\documentclass[%
preprint,
 amsmath,amssymb,
 aps,
]{revtex4-2}

\usepackage{graphicx}
\usepackage{dcolumn}
\usepackage{bm}

\begin{document}


\title{Spatiotemporal Patterns Corresponding to Phase Synchronization and Generalized Synchronization States of Thermoacoustic Instability}

\author{Samadhan A. Pawar}
\email{samadhanpawar@ymail.com}
 \affiliation{Institute for Aerospace Studies, University of Toronto, Toronto, ON M3H5T6, Canada}

\author{P. R. Midhun}%
\affiliation{ Department of Aerospace Engineering, Indian Institute of Technology Madras, Chennai, TN 600036, India.}%

\author{K. V. Reeja}%
\affiliation{ Department of Aerospace Engineering, Indian Institute of Technology Madras, Chennai TN 600036, India.}%

\author{Abin Krishnan}%
\affiliation{ Daniel Guggenheim School of Aerospace Engineering, Georgia Institute of Technology, Atlanta, GA 30332, USA}%

\author{Krishna Manoj}%
\affiliation{ Department of Mechanical Engineering, Massachusetts Institute of Technology, Cambridge, MA 02139, USA}%

\author{R. I. Sujith}%
\affiliation{ Department of Aerospace Engineering, Indian Institute of Technology Madras, Chennai, TN 600036, India}%


\date{\today}

\begin{abstract}
Thermoacoustic instability in turbulent combustion systems emerges from the complex interplay among the flame, flow, and acoustic subsystems. While the onset of thermoacoustic instability exhibits global order in system dynamics, the characteristics of local interactions between subsystems responsible for this order are not well understood. In this study, we utilize the framework of synchronization to elucidate the spatiotemporal interactions among heat release rate fluctuations in the flame, velocity fluctuations in the flow, and acoustic pressure fluctuations in a turbulent combustor. We examine two forms of thermoacoustic instability, characterized by phase synchronization and generalized synchronization of the acoustic pressure and global heat release rate oscillations. Despite the presence of global synchrony, we uncover a coexistence of frequency synchrony and desynchrony in the local interaction of these oscillations within the reaction field. In regions of frequency-locked oscillations, various phase-locking patterns occur, including phase synchrony and partial phase synchrony. We discover that the local development of small pockets of phase synchrony and strong amplitude correlation between these oscillations is sufficient to trigger global phase synchronization in the system. Furthermore, as the global dynamics approach generalized synchronization, these local regions of synchrony expand throughout the reaction field. Additionally, through coupled analysis of acoustic pressure and local flow velocity fluctuations, we infer that the spatial region of flow-acoustic synchrony plays a significant role in governing thermoacoustic instabilities. Our findings imply that, in turbulent combustors, an intrinsic local balance between order, partial order, and disorder within the coupled subsystems sustains the global order during thermoacoustic instability.  
\end{abstract}

\maketitle


\section{Introduction}

Over the years, the generation of self-sustained, high amplitude sound waves in confined turbulent combustion systems has been a serious obstacle to the development of gas turbines and rocket engines. These undesirable acoustic pressure oscillations are commonly known as thermoacoustic instability or combustion instability \citep{putnam1971combustion,lieuwen2005combustion,culick2006unsteady}. Thermoacoustic instabilities primarily arise due to positive feedback (mutual synchrony) between the acoustic field of the combustor and the unsteady heat release rate field of the flame. This physical explanation was first introduced by Rayleigh \cite{rayleigh1878explanation} and has subsequently been mathematically formulated by many researchers \citep{putnam1971combustion,chu1965energy,nicoud2005thermoacoustic}. Prolonged exposure to such instabilities can lead to several consequences for the structural, mechanical, and electrical components of the engine. Increased heat transfer during such instabilities can induce thermal stresses, leading to fatigue failure of components such as combustor liners, fuel injectors, and turbine blades \citep{lieuwen2005combustion}. Additionally, thermoacoustic instabilities can cause other issues, such as compressor surge, flame flashback, and blowout \citep{lieuwen2005combustion,lieuwen2012unsteady}. As a result, it is important to understand the characteristics of these instabilities and develop controls to protect combustion systems from them.  

Turbulent combustion systems by nature are spatially extended complex systems, wherein the interactions within the system are influenced by the spatial arrangement of its components \citep{sujith2021thermoacoustic}. Thermoacoustic instability in such systems results from the nonlinear interaction of various subsystems within the combustor \citep{lieuwen2012unsteady,poinsot2005theoretical,candel2002combustion,ducruix2003combustion,polifke2004combustion,oconnor2022understanding}. These include the turbulent flow field, reaction field, and chamber acoustics, interacting over a wide range of temporal and spatial scales. In recent years, various methodologies from nonlinear dynamics and complex system theory have been employed to better understand the onset of thermoacoustic instabilities and the different features of flow-flame-acoustic coupling present in these systems \citep{sujith2021thermoacoustic,zhao2023thermoacoustic}. In many turbulent thermoacoustic systems, the transition from a state of stable operation (combustion noise) to a state of unstable operation (thermoacoustic instability) occurs via an intermediate state of intermittency \citep{nair2014intermittency}. During this intermittency route to thermoacoustic instability, the dynamical properties of acoustic pressure change from chaotic oscillations to limit cycle oscillations through intermittent oscillations that consist of bursts of periodicity interspersed with epochs of aperiodicity \citep{sujith2021thermoacoustic}. Furthermore, depending on the characteristics of thermoacoustic instability, this state has been described as noisy limit cycle oscillations \citep{lieuwen2002experimental,noiray2013deterministic} or pseudoperiodic oscillations \citep{okuno2015dynamics} in turbulent combustors.

The spatial field of turbulent combustors exhibits intriguing patterns during the limit cycle state of thermoacoustic instability \citep{sujith2021thermoacoustic}. The oscillations in the CH* chemiluminescence field of the flame are primarily governed by underlying hydrodynamic fluctuations, vortex formation, equivalence ratio fluctuations, flame stabilization mechanics, and the boundaries of the combustor \citep{lieuwen2012unsteady}. Regardless of the combustor configuration and flow conditions, previous studies have demonstrated the periodic emergence of large-scale vortices along the shear layer in the combustor, leading to periodic modulation in the flame surface area and consequent heat release rate fluctuations during thermoacoustic instability \citep{rogers1956mechanism,smith1985combustion,poinsot1987vortex,schadow1992combustion,renard2000dynamics,shanbhogue2009flame,chakravarthy2007experimental,oconnor2022understanding}. This synchronized locking of flow, flame, and acoustic fields is sometimes referred to as vortex-acoustic lock-in  \citep{chakravarthy2007experimental,singh2019experimental,chakravarthy2017dynamics,matveev2003model,seshadri2016reduced}. By closely analyzing the vortex formation process, George \textit{et al.} \cite{george2018pattern} observed collective behavior among small-scale and large-scale vortices, where the interaction among small-scale vortices aids in building up a large-scale vortex in the system during each acoustic cycle.  

Despite extensive research on understanding the overall temporal and spatiotemporal dynamics of thermoacoustic instability using various perspectives, there is still limited knowledge about how the acoustic field of the combustor and turbulent reacting flow field interact locally during this state. In this paper, we pose several crucial questions that arise in this context: what happens to the turbulent reacting flow, which is typically chaotic and disordered during stable operation, during the occurrence of thermoacoustic instability? Is there a perfect order throughout the turbulent reaction field during this state, as observed in the global dynamics of the system? Furthermore, how does the spatiotemporal coupling between flow, flame, and acoustic fluctuations change when the temporal properties of limit cycle oscillations shift from weakly correlated to strongly correlated oscillations? To address these questions, we utilize the framework of synchronization theory \citep{pikovsky2003synchronization,boccaletti2018synchronization} and collective behaviour \citep{sumpter2010collective,satz2020rules} to provide a systematic investigation.  

The phenomenon of synchronization refers to the matching of the rhythms, both in terms of phase and frequency, of two or more oscillators due to the presence of mutual coupling between them \citep{pikovsky2003synchronization,strogatz2004sync,balanov2009synchronization}. In essence, synchronization leads to an invariant phase difference between the coupled oscillators over time, while collective behavior refers to the global coordinated behavior of a large group of individuals that emerge from the local interactions between the individuals and their environment \citep{sumpter2010collective}. Collective behavior is evident in flocks of birds, flashing of light by fireflies, and ant colonies, among others. In many cases, collective behavior arises as a result of synchronization between the individual elements of a group. Therefore, synchronization is a local phenomenon, while collective behavior is a global phenomenon that occurs in a large population of oscillators. 

 In recent thermoacoustic literature, various researchers have adopted the framework of mutual synchronization to investigate different aspects of thermoacoustic instability   \citep{pawar2017thermoacoustic,mondal2017onset,mondal2017synchronous,chiocchini2017chaotic,murayama2019attenuation,moeck2019nonlinear,guan2019chaos,pawar2019temporal,ramanan2022detection,kasthuri2022coupled,dutta2019investigating,bonciolini2019synchronization,hashimoto2019spatiotemporal,guan2022synchronization,weng2023synchronization}. In a turbulent combustion system, the acoustic field in the combustor and the heat release rate fluctuations in the flame have been viewed as two coupled non-identical oscillators \citep{pawar2017thermoacoustic,mondal2017onset,mondal2017synchronous,kheirkhah2017non,chiocchini2017chaotic,godavarthi2018coupled,guan2019chaos,singh2023mean,weng2023synchronization}. In the absence of combustion, the low amplitude self-sustained chaotic oscillations are observed in the acoustic pressure due to the underlying turbulent flow. Conversely, the heat release rate fluctuations in open turbulent flame (i.e., in the absence of any acoustic perturbations due to confinement) exhibit self-sustained broadband aperiodic fluctuations. In a confined combustion system, these oscillators are weakly coupled during the state of combustion noise and are strongly coupled during the state of thermoacoustic instability \citep{pawar2017thermoacoustic,singh2023mean}. Both acoustic pressure and heat release rate fluctuations display self-sustained limit cycle oscillations during thermoacoustic instability \citep{mondal2017synchronous}. Even though both the acoustic field and turbulent reacting field coexist in the same system, their synchronization can be studied in a similar manner to understanding the collective firing of an ensemble of neurons in brain research \citep{schafer1999synchronization}, the coupling between heart rhythm and respiration in the human body \citep{uhlhaas2006neural}, and revealing variety of spatial patterns observed in the interaction of chemical oscillators \citep{taylor2009dynamical}.

 During the state of thermoacoustic instability, the dominant frequencies of acoustic pressure and heat release rate fluctuations are perfectly locked. However, their instantaneous phase interactions can exhibit a variety of patterns. By analyzing the synchronization properties of these fluctuations in a turbulent combustor, Pawar \textit{et al.} \cite{pawar2017thermoacoustic} characterized the state of thermoacoustic instability into phase synchronization and generalized synchronization. Similar observations have also been reported in different combustion systems \citep{pawar2018synchronization,guan2019chaos,kasthuri2022coupled}. Recently, Kushwaha \textit{et al.} \cite{kushwaha2021dynamical} observed an interesting 2:1 phase-locking between acoustic pressure and heat release rate fluctuations in a hydrogen-enriched methane combustion system. Here, two cycles of acoustic pressure fluctuations are locked with a single cycle of heat release rate oscillations during period-2 thermoacoustic instability. 
 Mondal \textit{et al.} \cite{mondal2017onset} used the synchronization framework to explore the relationship between acoustic pressure and local heat release rate oscillations during the transition to thermoacoustic instability. They found that the phasor field of these coupled oscillations evolved from disorder to order via a state of chimera, where both order and disorder coexisted. Similarly, Dutta \textit{et al.} \cite{dutta2019investigating} and Singh \textit{et al.} \cite{singh2023mean} modeled the phase interaction between acoustic pressure and local heat release rate fluctuations using the Kuramoto phase oscillator model. Furthermore, researchers have employed the mutual synchronization approach to investigate the interaction between multiple turbulent flames and the acoustic field in an annular combustor \citep{roy2020flame} and a model liquid rocket combustor \citep{kasthuri2022coupled}, and the interaction of acoustic fields in an annular combustor \citep{moeck2019nonlinear} and multiple can combustors   \citep{farisco2017thermo, jegal2019mutual,guan2022synchronization,moon2023modal}, during the state of thermoacoustic instability.

To investigate the coupling between flow, flame, and acoustics during thermoacoustic instability, we perform experiments on a bluff-body stabilized turbulent combustor, by varying the global equivalence ratio (or Reynolds number of the airflow) as a system parameter. We found that the route to thermoacoustic instability follows the transition: combustion noise $\rightarrow$ intermittency $\rightarrow$ weakly correlated limit cycle $\rightarrow$ strongly correlated limit cycle. These two types of limit cycle oscillations represent distinct forms of thermoacoustic instability. By analyzing the interaction between acoustic pressure ($p^\prime$) and global heat release rate ($\dot{q}^\prime$) in the flame, we classify these oscillations as phase synchronization and generalized synchronization, respectively. Our analysis reveals that during the transition from phase synchronization to generalized synchronization, the variation in phase, frequency, and amplitude of $p^\prime$ and global $\dot{q}^\prime$ oscillations are strongly correlated. We also examine the spatial coupling between $p^\prime$ and local $\dot{q}^\prime$ oscillations using measures such as spectral coherence, phase locking value, and Pearson correlation coefficient. Our results show that although the global oscillations in $p^\prime$ and $\dot{q}^\prime$ are perfectly synchronized and ordered during both states of thermoacoustic instability, their local coupling exhibits complex dynamics ranging from desynchrony to partial synchrony to phase synchrony. In the phase synchronization regime of thermoacoustic instability, we observe that small pockets of local phase synchrony and a strong correlation between $p^\prime$ and local $\dot{q}^\prime$ oscillations emerge close to the combustor wall. These regions then expand in size and spread across the reaction field during the occurrence of generalized synchronization. Finally, we investigate the local coupling between acoustic pressure and flow velocity fluctuations during these states and identify critical regions in the reaction field. Using secondary micro-jet injection of the airflow, we compare the suppression characteristics of thermoacoustic instability during the states of phase synchronization and generalized synchronization. 

The paper is organized as follows: In Sec. II, we provide a detailed description of the experimental setup and the different methods used for data acquisition. In Sec. III, we first discuss the distinction between phase synchronization and generalized synchronization states that occur in the acoustic pressure and global heat release rate oscillations during thermoacoustic instability. We then analyze the local coupling between these oscillations in the reaction field, examining their frequency, phase, and amplitude locking behaviors. Additionally, we investigate the local coupling between the acoustic pressure and velocity fluctuations in the system. Utilizing the synchronization properties of these oscillations, we identify critical regions in the reaction field and develop a methodology to control such instabilities. In Sec. IV, we summarize the key findings from our study and discuss the importance of studying turbulent combustion systems from a complex systems perspective.

\section{Experimental setup and data acquisition}

\begin{figure}
  \centerline{\includegraphics[scale=1.4]{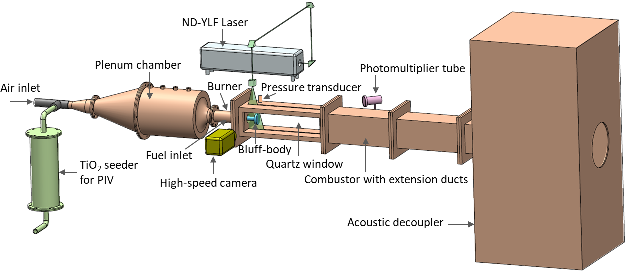}}
  \caption{Schematic of the experimental setup and the arrangement for optical flow diagnostics and measurements.}
\label{fig1}
\end{figure}

In Fig. \ref{fig1}, we present a schematic of a laboratory-scale bluff-body stabilized dump combustor with a partially premixed turbulent flame. The experimental setup includes a plenum, a burner, and a rectangular combustion chamber with extension ducts. The combustor consists of a backward-facing step (dump plane) at the inlet and a circular bluff-body (47 mm diameter and 10 mm thickness) for flame stabilization. A central shaft (diameter 16 mm) passes through the burner and is used to hold the bluff-body at a distance of 35 mm from the dump plane. The central shaft is also utilized to supply fuel into the combustor through four radial injection holes, each of 1.7 mm diameter, located 160 mm upstream of the rear end of the bluff-body. To prevent flashback, a flashback arrestor made of a 2 mm thick circular disk with 300 holes, each of 1.7 mm diameter, is positioned in the burner 30 mm downstream of the fuel-injection location. The combustion chamber has a cross-sectional area of 90$\times$90 mm$^2$ and a length of 1100 mm. Two quartz windows, with dimensions of 400$\times$90$\times$10 mm$^3$, are provided on opposite walls of the combustion chamber for optical diagnostics. 

Liquified petroleum gas (LPG) with a composition of 60\% butane and 40\% propane is used as fuel. The compressed air enters the combustor through a plenum, which reduces fluctuations in the inlet air flow. It is then partially premixed with the fuel in the burner section. A spark plug powered by a step-up transformer (11 kV) is mounted on the dump plane to ignite the partially premixed fuel-air mixture. The fuel and air mass flow rates are controlled using Alicat Scientific MCR series mass flow controllers with an uncertainty of $\pm$0.8\% and a full-scale uncertainty of 0.2\%. The Reynolds number ($Re$) of the unburnt fuel-air mixture at the burner is varied between 0.79$\times$10$^5$ and 1.36$\times$10$^5$, while the global equivalence ratio ($\Phi$) of this mixture in the combustor is varied from 0.97 to 0.52.

A smart control methodology that utilizes secondary air injection into the combustion chamber is employed to suppress thermoacoustic instabilities. The secondary air is introduced as micro-jets through three ports (indicated by P1, P2, and P2 in Fig. \ref{fig12}a) of 5 mm diameter mounted on the top and bottom walls of the combustor. The micro-jet is aimed at the critical region, which is determined through the local phase analysis of coupled flow velocity and acoustic pressure fluctuations in the reaction field. A separate mass flow controller is used to regulate the secondary air. During smart control experiments, we first establish a particular state of thermoacoustic instability in the combustor by controlling the primary air and fuel flow rates. Then, we increase  the flow rates of secondary air in small steps through a particular port and examine the suppression characteristics of thermoacoustic instability in the system.

A piezoelectric pressure transducer (PCB103B02) is utilized to acquire unsteady acoustic pressure fluctuations in the combustor, while a photomultiplier tube (Hamamatsu H10722-01) is used to capture global heat release rate fluctuations in the flame. The pressure transducer is mounted 25 mm away from the dump plane of the combustor, while the photomultiplier tube with a CH* filter (narrow bandwidth filter centered at 435 nm and 10 nm full width at half maximum FWHM) is positioned in front of the optical access of the combustor to detect the total chemiluminescence intensity in the flame. An analog-to-digital card (NI-6143, 16-bit) is utilized to record both the acoustic pressure and global heat release rate signals for 3 s at a sampling rate of 13.5 kHz. Additionally, a high-speed CMOS (Photron FASTCAM SA4) camera and a ZEISS 100 mm lens with an aperture at $f/2$ are used to capture the simultaneous CH* chemiluminescence field of the flame (camera resolution: 1200$\times$600 pixels). The images are recorded for 0.46 s at a sampling rate of 2700 frames per second.

To obtain the local velocity field of the turbulent reactive flow, we performed high-speed two-component particle image velocimetry (PIV) measurements, in conjunction with simultaneous measurements of acoustic pressure, global heat release rate, and high-speed CH* chemiluminescence images of the flame. A single-cavity double-pulsed Nd:YLF laser (Photonics) with a 527 nm wavelength, operating at a repetition rate of 2 kHz and a pulse duration of 250 ns, is used to produce twin pulses for illuminating the seeding particles. The laser beam is directed toward the combustion chamber using a set of right-angled prisms and convex lenses (500 mm and 50 mm). This beam is then expanded into a 2 mm thick laser sheet by passing it through 600 mm spherical and –1 mm cylindrical lenses. We utilized TiO2 particles of approximate size 1 µm to seed the flow. The separation time between the two pulses was carefully chosen, with a time delay of 15 to 25 µs selected depending on the mean velocity of the flow, to ensure that the maximum pixel displacement of the particles falls within the range of approximately 4 to 7 pixels between the two laser pulses. We also ensured that the maximum particle displacement does not exceed 1/4$^{th}$ of the size of the interrogation window to avoid the in-plane losses of particles during the PIV evaluation.

The scattered light from the seeding particles (Mie scattering) is captured as a sequence of frames using a high-speed CMOS camera (Photron FASTCAM SA4) synchronized with the laser. The camera with a maximum resolution of 1024$\times$1024 pixels is equipped with a ZEISS 100 mm focal length lens at $f/5.6$ aperture. Image pairs are captured at a rate of 2000 Hz. For our experiments, a measurement region of 50$\times$36 mm is imaged onto 1024$\times$736 pixels of the sensor, which corresponds to the upper half of the bluff body where PIV is performed. To capture the Mie scattering light, a short bandpass optical filter centered at 527 nm (12 nm FWHM) was mounted in front of the lens. The post-processing analysis of the PIV data is discussed in detail in Unni \textit{et al.} \cite{unni2018emergence}. 


\section{Results and discussion}
\subsection{Dynamics during the transition to thermoacoustic instability}

\begin{figure}
  \centerline{\includegraphics[scale=1.4]{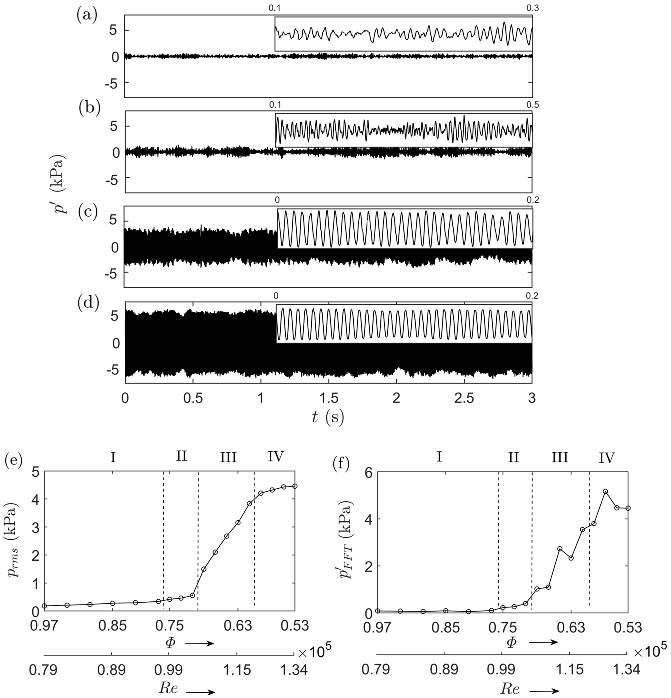}}
  \caption{The time series of acoustic pressure fluctuations ($p^\prime$) are shown for the states of (a) stable operation (i.e., high-dimensional chaos), (b) intermittency, (c) weakly correlated, and (d) strongly correlated limit cycle oscillations, as indicated by regions I, II, III, and IV, respectively, in (e) and (f). An inset shows the zoomed-in portion of the $p^\prime$ signal for a corresponding state. The global equivalence ratio ($\Phi$) values for these states are 0.93, 0.71, 0.67, and 0.53, respectively. The bifurcation diagrams demonstrate how (e) the root mean square value of $p^\prime$ fluctuations and (f) the amplitude of the dominant frequency change with decreasing $\Phi$ (or increasing Reynolds number of the airflow $Re$), as the system transitions from chaos to strongly correlated limit cycle oscillations.}
\label{fig2}
\end{figure}

Most often, thermoacoustic instabilities are characterized by analyzing the temporal properties of acoustic pressure in the combustor. In Figs. \ref{fig2}a to \ref{fig2}d, we demonstrate the change in the dynamic behavior of acoustic pressure fluctuations ($p^\prime$) when the global equivalence ratio is varied as a control parameter, during the transition from combustion noise to thermoacoustic instability. As the system undergoes this dynamical transition, different regimes of acoustic pressure oscillations can be observed. When the global equivalence ratio is near the stoichiometric limit (Fig. \ref{fig2}a), low amplitude chaotic oscillations \citep{nair2013loss} are observed in the $p^\prime$ signal. This state is commonly referred to as stable operation or combustion noise \citep{dowling2015combustion}. As the global equivalence ratio is decreased towards a fuel-lean condition (Fig. \ref{fig2}b), the system behavior exhibits intermittent fluctuations in the $p^\prime$ signal. This intermittency state is characterized by the occurrence of bursts of periodic oscillations amidst regions of low amplitude aperiodic oscillations, as reported by Nair \textit{et al.} \cite{nair2014intermittency}. The amplitude of these bursts grows and the duration of their occurrence in the $p^\prime$ signal increases as the global equivalence ratio is decreased in the intermittency regime.

With further reduction in the global equivalence ratio, the system undergoes a transition from the state of intermittency to limit cycle oscillations (LCO), which are commonly known as thermoacoustic instability \citep{awad1986existence,ananthkrishnan2005reduced, lieuwen2002experimental,noiray2013deterministic}. We observe two forms of LCO in the system, namely weakly correlated (Fig. \ref{fig2}c) and strongly correlated (Fig. \ref{fig2}d) limit cycle oscillations \citep{pawar2017thermoacoustic}. These terms are based on the distinct properties of the coupled $p^\prime$ and global $\dot{q}^\prime$ oscillations observed during these states in the system, which are further elaborated in Figs. \ref{fig3} and \ref{fig4}.
 
To illustrate the transition from combustion noise to strongly correlated limit cycle oscillations, we present bifurcation diagrams in Fig. \ref{fig2}e and \ref{fig2}f, which depict the variation of the root mean square value ($p_{rms}$) of the $p^\prime$ signal and the magnitude of the dominant frequency peak ($p_{FFT}$) in the $p^\prime$ spectrum, respectively. The different dynamical behaviors of the combustor in the bifurcation diagram are characterized by examining both the temporal properties of $p^\prime$ fluctuations (Figs. \ref{fig2}a-d) and the coupled behavior of $p^\prime$ and global $\dot{q}^\prime$ fluctuations (as discussed in Fig. \ref{fig4}). We observe that the amplitude of $p^\prime$ oscillations remains consistently low during regimes of stable operation and intermittency. However, during the onset of weakly correlated limit cycle oscillations, the amplitude of $p^\prime$ oscillations experiences a sudden increase. This increase continues until it nearly saturates at a higher value during the occurrence of strongly correlated limit cycle oscillations.

\subsection{Temporal properties of phase synchronization and generalized synchronization}

Having discussed how the temporal properties of $p^\prime$ oscillations change during different states, we next focus on investigating the dynamical properties of thermoacoustic instability alone. In particular, we present how both temporal and spatiotemporal properties of coupled $p^\prime$ and $\dot{q}^\prime$ oscillations change as the system behavior transitions from the state of weakly correlated LCO to strongly correlated LCO. Pawar \textit{et al.} \cite{pawar2017thermoacoustic} previously characterized the states of weakly correlated LCO and strongly correlated LCO in the coupled $p^\prime$ and global $\dot{q}^\prime$ fluctuations as phase synchronization and generalized synchronization, respectively, by studying their recurrence properties. During phase synchronization, the instantaneous phases of the coupled oscillators are perfectly locked in time, while their amplitudes show a low correlation, indicating a weak form of synchronization \citep{rosenblum1996phase}. Conversely, during generalized synchronization, perfect locking of instantaneous phases and amplitudes is observed, indicating that the dynamics of oscillators are locked in a particular functional relationship, and the system displays a strong form of synchronization \citep{rulkov1995generalized}. We will use the terms phase synchronization and generalized synchronization to refer to the weakly and strongly correlated LCO states in the $p^\prime$ and global $\dot{q}^\prime$ oscillations, respectively, throughout the rest of our paper. 

\begin{figure}
  \centerline{\includegraphics[scale=0.4]{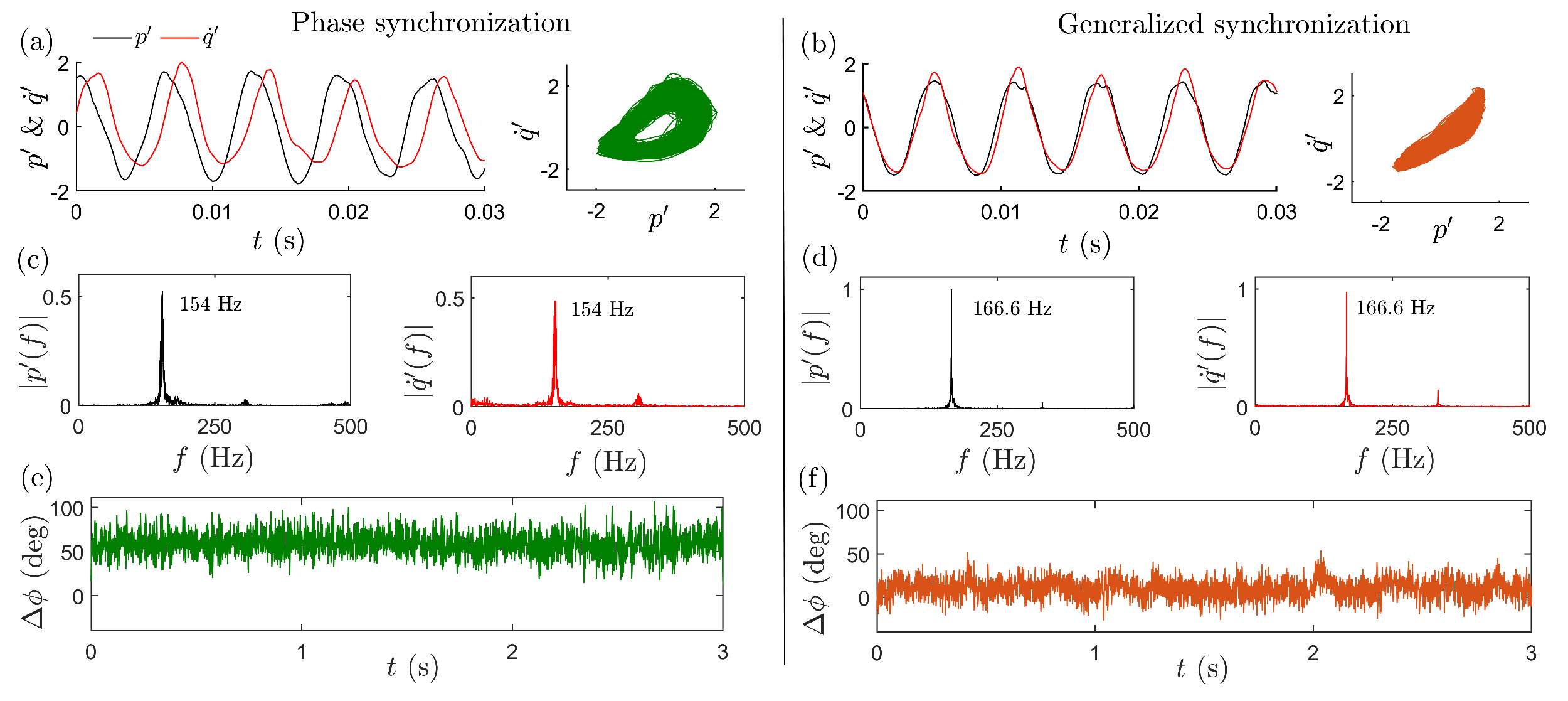}}
  \caption{(a, b) The normalized $p^\prime$ (in black) and global $\dot{q}^\prime$ (in red) signals along with the corresponding phase portrait obtained by plotting $p^\prime$ versus $\dot{q}^\prime$, (c, d) the amplitude spectrum of these signals, and (e, f) the time series of the relative phase ($\Delta\phi$) between them, for the state of phase synchronization and generalized synchronization during thermoacoustic instability, respectively. We observe that $p^\prime$ and global $\dot{q}^\prime$ oscillations are frequency and phase-locked during both states of thermoacoustic instability; however, they exhibit different correlation values and mean phase differences during these states, as discussed in Fig. \ref{fig4}.}
\label{fig3}
\end{figure}

Figures \ref{fig3}c and \ref{fig3}d depict the amplitude spectrum of $p^\prime$ and global $\dot{q}^\prime$ oscillations (Figs. \ref{fig3}a,b) during the states of phase synchronization and generalized synchronization, respectively. We notice that the dominant frequencies of both $p^\prime$ and global $\dot{q}^\prime$ oscillations are perfectly locked within each state, indicating frequency synchronization. Interestingly, the dominant frequency is slightly higher during the state of generalized synchronization (166.6 Hz) compared to the state of phase synchronization (154 Hz) in our system. 

Next, we obtain the instantaneous phase of a signal using the analytic signal approach, which involves applying the Hilbert transform to the signal \citep{pikovsky2003synchronization}. The analytic signal of $p^\prime(t)$ is represented as follows: 
\begin{equation}
\zeta_{p^\prime} (t) = p^\prime(t) + i \mathrm{HT}[p^\prime(t)]
\label{eq1}
\end{equation}
where $i = \sqrt{-1}$ and HT is the Hilbert transform of the signal obtained from the following equation:
\begin{equation}
\mathrm{HT}[p^\prime(t)]=P.V.\int_{-\infty}^{\infty}\frac{p^\prime(\tau)}{t-\tau} d\tau.
\label{eq2}
\end{equation}
Here, $P.V.$ represents the Cauchy principal value. Since $\zeta_{p^\prime}(t)$ is a complex signal, we obtain the instantaneous amplitude and phase of the signal using the following expressions:
\begin{equation}
\zeta_{p^\prime}(t) = A_{p^\prime}(t) e^{i\phi_{p^\prime} (t)},	
\label{eq3}
\end{equation}
where $A_{p^\prime}(t) = \sqrt{[p^\prime(t)]^2 + [\mathrm{HT}(p^\prime(t))]^2}$ and $\phi_{p^\prime}(t) = \tan^{-1} [{\mathrm{HT}(p^\prime(t))}/{p^\prime(t)]}$.

A similar analysis can be performed for $\dot{q}^\prime(t)$ signal. Finally, the instantaneous phase difference between $p^\prime$ and $\dot{q}^\prime$ signals is obtained as:
\begin{equation}
\Delta\phi(t) = \phi_{p^\prime}(t) - \phi_{\dot{q}^\prime} (t)
\label{eq4}
\end{equation}
In the synchronized state, the relative phase between the oscillators remains constant over time, whereas in the desynchronized state, the relative phase drifts without bounds. In the intermittent synchronization state, the relative phase remains constant during the synchronized epochs of oscillations but undergoes phase slips in multiples of $2\pi$ radians during the desynchronized epochs. 

In Figs. \ref{fig3}e and \ref{fig3}f, we show the temporal variation of the unwrapped instantaneous phase difference ($\Delta\phi$) between $p^\prime$ and $\dot{q}^\prime$ signals for the state of phase synchronization and generalized synchronization, respectively. In both states, $\Delta\phi$ remains nearly constant around a mean value. However, the mean phase difference is relatively higher (around 59.5$^\circ$) for the phase synchronization state (Fig. \ref{fig3}e), while it is closer to 0$^\circ$ (around 10.7$^\circ$) for the generalized synchronization state (Fig. \ref{fig3}f). 

To examine the amplitude correlation between $p^\prime$ and $\dot{q}^\prime$ signals, we plot these signals against each other. Since these signals are periodic, the behavior of phase space trajectories repeats after a time period of the oscillations (Fig. \ref{fig3}a,b). During the state of phase synchronization, where $p^\prime$ and $\dot{q}^\prime$ signals are locked at a phase difference much greater than zero, we observe the presence of a distorted elliptic-shaped attractor in the phase space (Fig. \ref{fig3}a). These distortions in the phase space occur due to the wide cycle-to-cycle variation in the amplitude of periodic oscillations observed during this state \citep{pawar2017thermoacoustic}. In contrast, for the state of generalized synchronization (Fig. \ref{fig3}b), where the phase difference between $p^\prime$ and $\dot{q}^\prime$ signals is close to zero, the trajectories appear to be almost overlapped in the phase space.  

Over the years, several methods have been proposed to investigate the characteristics of phase synchronization and generalized synchronization of coupled $p^\prime$ and $\dot{q}^\prime$ oscillations in turbulent thermoacoustic systems. These methods include the interdependence index \citep{chiocchini2017chaotic}, order parameter \citep{mondal2017onset,murayama2019attenuation}, the probability of recurrence of the phase space trajectory \citep{pawar2017thermoacoustic}, cross wavelet transform \citep{pawar2018synchronization}, various measures from recurrence plots and recurrence networks \citep{godavarthi2018coupled}, and mutual information of two signals \citep{ghosh2022anticipating}. In this study, we employ the phase locking value (PLV) and Pearson correlation coefficient ($R$) to quantify the synchronization properties of $p^\prime$ and $\dot{q}^\prime$ oscillations \citep{mondal2017synchronous} observed during these states. These measures have been widely used in neurology to assess the complex interactions between neurons in the brain \citep{lachaux1999measuring,jalili2013synchronization,haken2006brain,liang2016comparison}.

The phase locking value (PLV) is a quantitative measure used to determine the synchronization of phases of two oscillators, which is obtained by calculating the following equation \citep{lachaux1999measuring}:
\begin{equation}
\mathrm{PLV}= \frac{1}{N}\Big|\sum_{j=1}^N e^{i\Delta \phi_j(t)}\Big|	
\label{eq5}
\end{equation}
Here, $N$ represents the total number of data points in the signal. The PLV ranges between 0 to 1, where a value near 1 indicates synchronized oscillations and a value near 0 indicates desynchronized oscillations.

\begin{figure}
  \centerline{\includegraphics[scale=0.6]{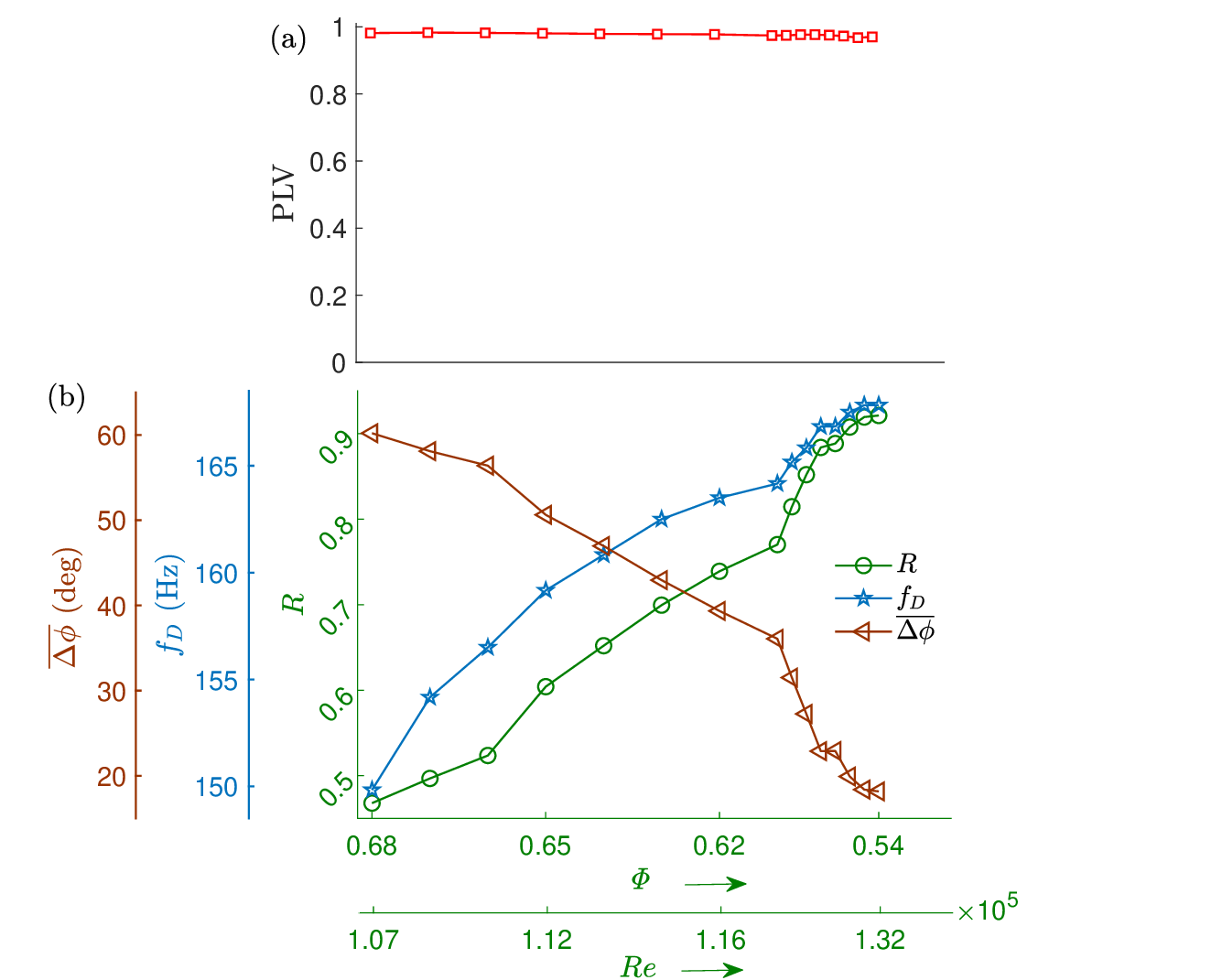}}
  \caption{(a) The variation of PLV and (b) $R$,  $\overline{\Delta\phi}$, and $f_D$ with respect to a decrease in the value of the global equivalence ratio ($\Phi$) as the system behavior transitions from the states of phase synchronization to generalized synchronization during thermoacoustic instability. During this transition, we notice that the PLV remains closer to unity, but the values of $R$,  $\overline{\Delta\phi}$, and $f_D$ exhibit a correlated variation. A combined analysis of these measures can aid in identifying the occurrence of the states of phase synchronization and generalized synchronization in the system.}
\label{fig4}
\end{figure}

Similarly, the Pearson correlation coefficient ($R$) can be used to determine the strength of amplitude synchronization between two oscillators \citep{gonzalez2002amplitude,bove2004frequency}. This measure provides information about how two continuous signals co-vary over time. $R$ is a linear measure with a value ranging from $-1$ (negatively correlated) to $0$ (not correlated) to $1$ (perfectly correlated). The Pearson correlation coefficient between $p^\prime$ and $\dot{q}^\prime$ signals is given by the following equation:
\begin{equation}
R = \frac{\sum_{j}(p_j^\prime - \overline{p^\prime})(\dot{q}_{j}^\prime - \overline{\dot{q}^\prime})}{\sqrt{\sum_{j}(p_j^\prime - \overline{p^\prime})^2 \sum_{j}(\dot{q}_{j}^\prime - \overline{\dot{q}^\prime})^2}}
\label{eq6}
\end{equation}
Here, the overbar indicates the temporal mean of the signal, and the summation is performed over the entire duration of the signal such that $j=1$ to $N$.  

It is worth noting that the expression for the Pearson correlation coefficient ($R$) is identical to that of the normalized Rayleigh index, which is commonly used to quantify the driving and damping regions of acoustic power production in the reaction field of a combustor \citep{putnam1971combustion}. The Rayleigh index is obtained by integrating the product of $p^\prime$ and $\dot{q}^\prime$ fluctuations measured for the combustor volume over one oscillation cycle, and can be expressed as follows: 
\begin{equation}
RI=\frac{1}{T}\int_0^T\int_V\tilde{p^\prime}\tilde{\dot{q}^\prime}dvdt
\label{eq7}
\end{equation}
where $\tilde{p^\prime}$ and $\tilde{\dot{q}^\prime}$ represent the normalized acoustic pressure and the heat release rate oscillations by their standard deviation, respectively, while $T$ is the time period of $p^\prime$ oscillations and $V$ is the volume of the reaction zone. The value of the normalized Rayleigh index is expected to be positive for the underlying heat release fluctuations in the flame to drive the acoustic field in the system.

Figure \ref{fig4} illustrates the variation in different measures quantifying the coupled behavior between $p^\prime$ and global $\dot{q}^\prime$ oscillations during the transition from phase synchronization to generalized synchronization in thermoacoustic instability. To investigate this transition, we systematically increase the mass flow rate of air while keeping the mass flow rate of fuel constant, resulting in a decrease in the global equivalence ratio ($\Phi$) of the reacting mixture towards a fuel lean condition. This variation in air flow rate leads to an increase in the Reynolds number ($Re$) of the flow, subsequently increase in the frequency of vortex shedding from the dump plane of the combustor.   During the state of thermoacoustic instability, the dominant frequencies of vortex shedding and acoustic pressure oscillations are perfectly locked (as shown in Figs. \ref{fig9} and \ref{fig10}). As a result, we notice a continuous rise in the dominant frequency ($f_D$) of $p^\prime$ oscillations from 149 to 167 Hz with an increase in $Re$, as depicted in Fig. \ref{fig4}b. Furthermore, as the instantaneous phases of $p^\prime$ and global $\dot{q}^\prime$ oscillations are  locked during the states of phase synchronization and generalized synchronization, PLV remains close to unity throughout the transition (Fig. \ref{fig4}a). However, during this transition, we observe a significant decrease in the value of mean phase difference ($\overline{\Delta\phi}$) between synchronized $p^\prime$ and global $\dot{q}^\prime$ oscillations from approximately 60° to 10°, as depicted in Fig. \ref{fig4}b. This is accompanied by an increase in the strength of amplitude correlation ($R$) between these oscillations from weak ($R$ close to 0.5) to strong ($R$ greater than 0.8) (Fig. \ref{fig4}b). 

These interrelated variations among $f_D$, $\overline{\Delta\phi}$, and $R$ suggest that as the natural frequency of $p^\prime$ and $\dot{q}^\prime$ oscillations increases, these oscillations progressively align their phases to lock near zero degrees of phase difference during the transition from phase synchronization and generalized synchronization. At this condition, heat addition occurs near the local peak of the acoustic pressure cycle, leading to stronger acoustic oscillations during the state of generalized synchronization compared to the state of phase synchronization. Furthermore, we can use PLV and $R$ between $p^\prime$ and $\dot{q}^\prime$ oscillations to accurately identify the onset of these states of synchronization in a thermoacoustic system. During the onset of phase synchronization, PLV will approach 1, and $R$ will exhibit a lower positive value. In contrast, during generalized synchronization, both PLV and $R$ will stay near unity.    


\subsection{Spatial patterns of flame-acoustic interaction: Frequency, phase and amplitude locking }

\begin{figure}
  \centerline{\includegraphics[scale=.9]{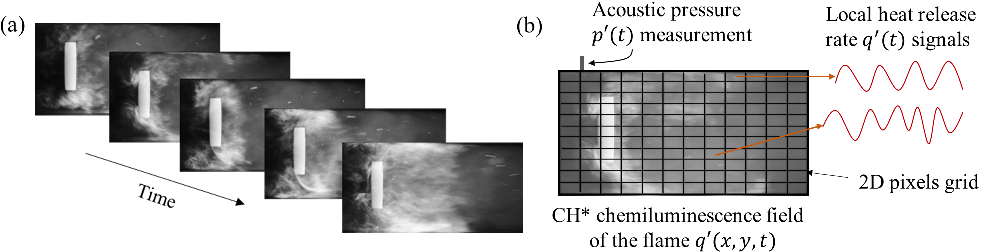}}
  \caption{(a) Sequential instantaneous images of CH* chemiluminescence field of the flame. (b) A schematic highlighting the local distribution of heat release rate fluctuations at a pixel (indicated by a grid) of the flame image acquired at time $t$.}
\label{fig5}
\end{figure}

After analyzing the temporal behavior of synchronized $p^\prime$ and global $\dot{q}^\prime$ oscillations during the transition from phase synchronization to generalized synchronization, we now investigate the spatiotemporal synchrony between $p^\prime$ and local $\dot{q}^\prime$ oscillations during these states of thermoacoustic instability. The spatial distribution of the line-of-sight integrated local CH* chemiluminescence field of the reaction zone is captured by high-speed imaging of the flame, as shown in Fig. \ref{fig5}a. Here, the local heat release rate fluctuations ($\dot{q}^\prime$) are obtained by discretizing the instantaneous spatial heat release rate information of the flame into pixels of an image, as indicated in Fig. \ref{fig5}b. Due to measurement limitations, acoustic pressure fluctuations are measured at a point close to the bluff-body where the flame is stabilized. Additionally, since the length of the reaction zone is much smaller than the wavelength of the acoustic waves generated during thermoacoustic instability in the combustor, the local variation of the magnitude of $p^\prime$ fluctuations is considered negligible inside the reaction zone \citep{mondal2017onset}. To understand the properties of local coupling between $p^\prime$ and $\dot{q}^\prime$ oscillations in the reaction field, we calculate the spatial distribution of various synchronization measures such as spectral coherence ($C$), PLV, and $R$. These measures help quantify the local synchrony in terms of frequency, phase, and amplitude of coupled oscillations, respectively. 

\begin{figure}
  \centerline{\includegraphics[scale=.83]{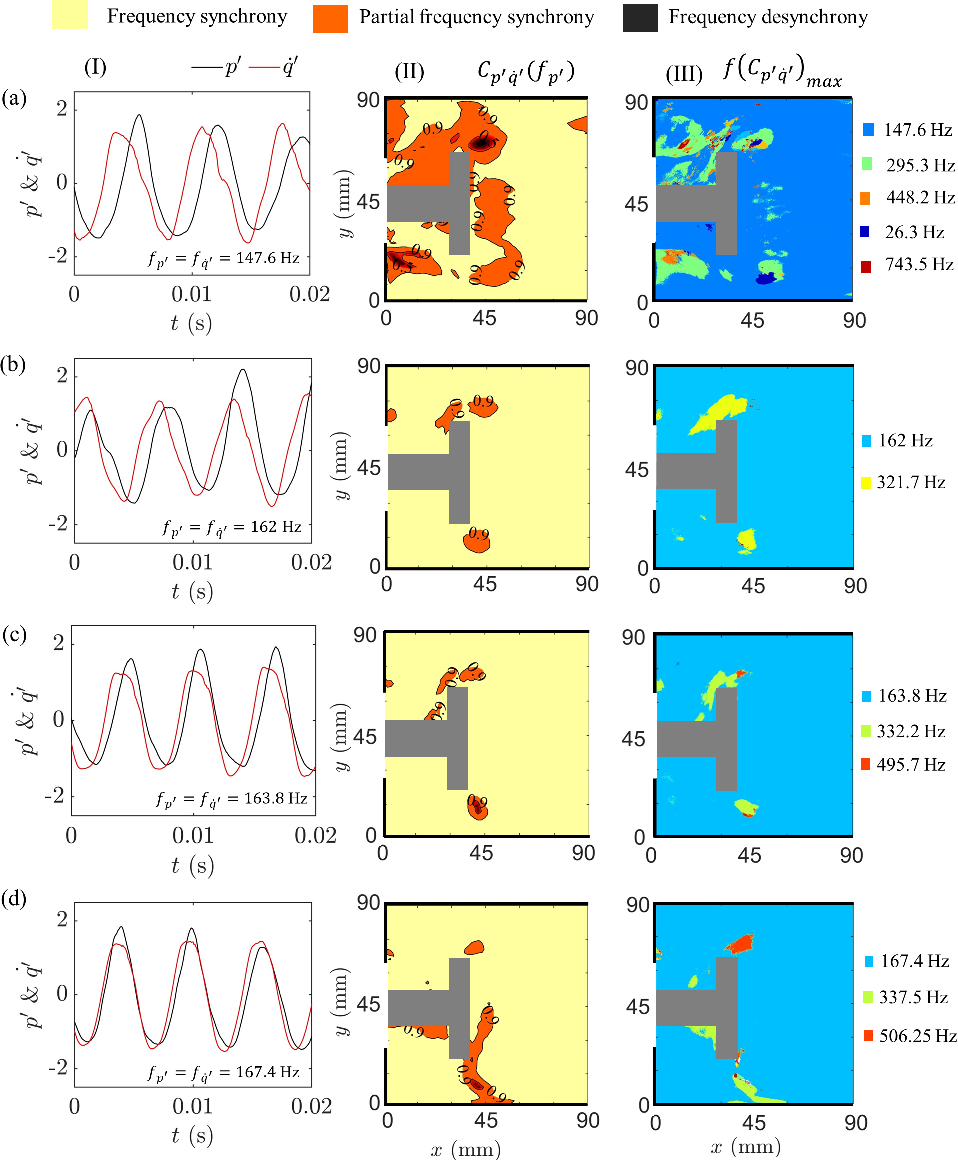}}
\caption{(I) The normalized time series of $p^\prime$ and global  $\dot{q}^\prime$ oscillations. (II) The spatial contours of spectral coherence $C_{p^\prime \dot{q}^\prime}$ between $p^\prime$ and local  $\dot{q}^\prime$ oscillations at the dominant frequency of acoustic pressure ($f_{p^\prime}$), highlighting regions of frequency synchrony (0.9 $\le$ $C_{p^\prime \dot{q}^\prime} (f_{p^\prime})$ $\le$ 1), partial frequency synchrony (0.5 $<$ $C_{p^\prime \dot{q}^\prime} (f_{p^\prime})$ $<$ 0.9), and frequency desynchrony (0 $\le$ $C_{p^\prime \dot{q}^\prime} (f_{p^\prime})$ $\le$ 0.5). (III) The spatial frequency clusters correspond to the maximum value of spectral coherence $C_{p^\prime \dot{q}^\prime}$. The synchronization frequency of $p^\prime$ and global $\dot{q}^\prime$ oscillations is indicated at the bottom right corner of each plot in (I). The plots in (a) to (d) are obtained for different states of thermoacoustic instability observed during the transition from phase synchronization to generalized synchronization, corresponding to the global equivalence ratio values of 0.68, 0.63, 0.6, and 0.54, respectively. The location of the bluff-body and shaft is masked in gray, while the combustor border is highlighted by black lines. We notice the coexistence of regions of frequency synchronized and desynchronized $p^\prime$ and local  $\dot{q}^\prime$ oscillations in the reaction field.}
\label{fig6}
\end{figure}

The spectral coherence, denoted as $C_{p^\prime \dot{q}^\prime} (f)$, quantifies the degree of linear dependence between the coupled $p^\prime$ and local  $\dot{q}^\prime$ oscillations at a specific frequency $f$ in the reaction field. Mathematically, it is defined as the ratio of the squared cross-power spectral density ($S_{p^\prime \dot{q}^\prime}$) between the $p^\prime$ and $\dot{q}^\prime$ signals and the product of their individual power spectral densities $S_{p^\prime p^\prime}$ and $S_{\dot{q}^\prime \dot{q}^\prime}$, as shown below \citep{lowet2016quantifying,carter1987coherence}:   
\begin{equation}
C_{p^\prime \dot{q}^\prime}(f)=\frac{|S_{p^\prime \dot{q}^\prime}(f)|^2}{S_{p^\prime p^\prime} (f) S_{\dot{q}^\prime \dot{q}^\prime}(f)}
\label{eq8}
\end{equation} 
The value of $C_{p^\prime \dot{q}^\prime} (f)$ ranges between 0 and 1, where 0 indicates the absence of linear dependency between $p^\prime$ and  $\dot{q}^\prime$ at $f$, while 1 represents the existence of a strong linear dependency between these oscillations at $f$. In other words, we can say that when the value of $C_{p^\prime \dot{q}^\prime} (f)$ is close to 1, the coupled $p^\prime$ and $\dot{q}^\prime$ oscillations are frequency-synchronized at $f$, whereas a value near 0 indicates their frequency desynchronization at $f$.  Thus, by analyzing the coherence spectrum across multiple frequencies, we can identify frequency bands where $p^\prime$ and $\dot{q}^\prime$ oscillators exhibit synchronization and frequency bands where synchronization between them is weak or absent. This information can provide valuable insights into the patterns of frequency-specific interaction among oscillators.

In Fig. \ref{fig6}, we present various characteristics of frequency locking between $p^\prime$ and $\dot{q}^\prime$ oscillations during different states of thermoacoustic instability, specifically focusing on the transition from phase synchronization to generalized synchronization. Figure \ref{fig6}-I demonstrates the time series of $p^\prime$ and global $\dot{q}^\prime$ oscillations, while Fig. \ref{fig6}-II shows the spatial distribution of $C_{p^\prime \dot{q}^\prime}$ calculated at the dominant frequency $f_{p^\prime}$ of $p^\prime$ oscillations. Based on the values of $C_{p^\prime \dot{q}^\prime}$ in the reaction field, the spatial distribution is categorized into different regions, named frequency desynchrony ($0 \le C_{p^\prime \dot{q}^\prime} (f_{p^\prime}) \le 0.5$), partial frequency synchrony ($0.5 < C_{p^\prime \dot{q}^\prime} (f_{p^\prime}) < 0.9$),  and frequency synchrony ($0.9 \le C_{p^\prime \dot{q}^\prime} (f_{p^\prime}) \le 1$), as indicated by different color codes at the top of Fig. \ref{fig6}. Additionally, Fig. \ref{fig6}-III displays the spatial distribution of frequency clusters corresponding to the maximum value of $C_{p^\prime \dot{q}^\prime}$, which are denoted by different colors on the right side of Fig. \ref{fig6}-III.

We observe that most of the spatial field of $C_{p^\prime \dot{q}^\prime} (f_{p^\prime})$ in Fig. \ref{fig6}-II exhibits values close to 1 across all states of thermoacoustic instability (Figs. \ref{fig6}a to \ref{fig6}d). This indicates that the broader region of local $\dot{q}^\prime$ oscillations in the reaction field is frequency-synchronized with dominant $p^\prime$ oscillations of the combustor. However, we also identify regions with lower coherence, indicated by partial frequency synchrony and frequency desynchrony, in the reaction field. Here, $p^\prime$ and local $\dot{q}^\prime$ oscillations are intermittently frequency synchronized during partial frequency synchrony, whereas they are completely uncorrelated at $f_{p^\prime}$ in the case of frequency desynchrony. The region of frequency-desynchronized oscillations is relatively small compared to that of partial frequency synchrony. These regions predominantly occur along the shear layers stabilized at the dump plane, shaft, and tip of the bluff-body. Notably, we find that the maximum coherence occurs at frequencies that are integer multiples of the dominant frequency $f_{p^\prime}$ of $p^\prime$ oscillations across all states (Figs. \ref{fig6}a to \ref{fig6}d). These frequencies are classified into different clusters, represented by various colors in Fig. 6-III. The presence of weak interaction between $p^\prime$ and local $\dot{q}^\prime$ oscillations during the onset of phase synchronization (Fig \ref{fig6}a-II) is evident from the higher number of frequency clusters observed in the reaction field as compared to other states (Fig \ref{fig6}b-II to Fig \ref{fig6}d-II). The frequency of the dominant cluster matches with the frequency of $p^\prime$ and global $\dot{q}^\prime$ oscillations shown in Fig 6-I. Thus, we observe a complex interplay in frequency interaction of $p^\prime$ and local $\dot{q}^\prime$ oscillations, where frequency-synchronized and desynchronized states coexist within the reaction field (Figs. \ref{fig6}-II, \ref{fig6}-III), despite the global behaviour of the system demonstrating perfect frequency-locking of these oscillations (Fig. \ref{fig6}-I).


\begin{figure}
  \centerline{\includegraphics[scale=0.8]{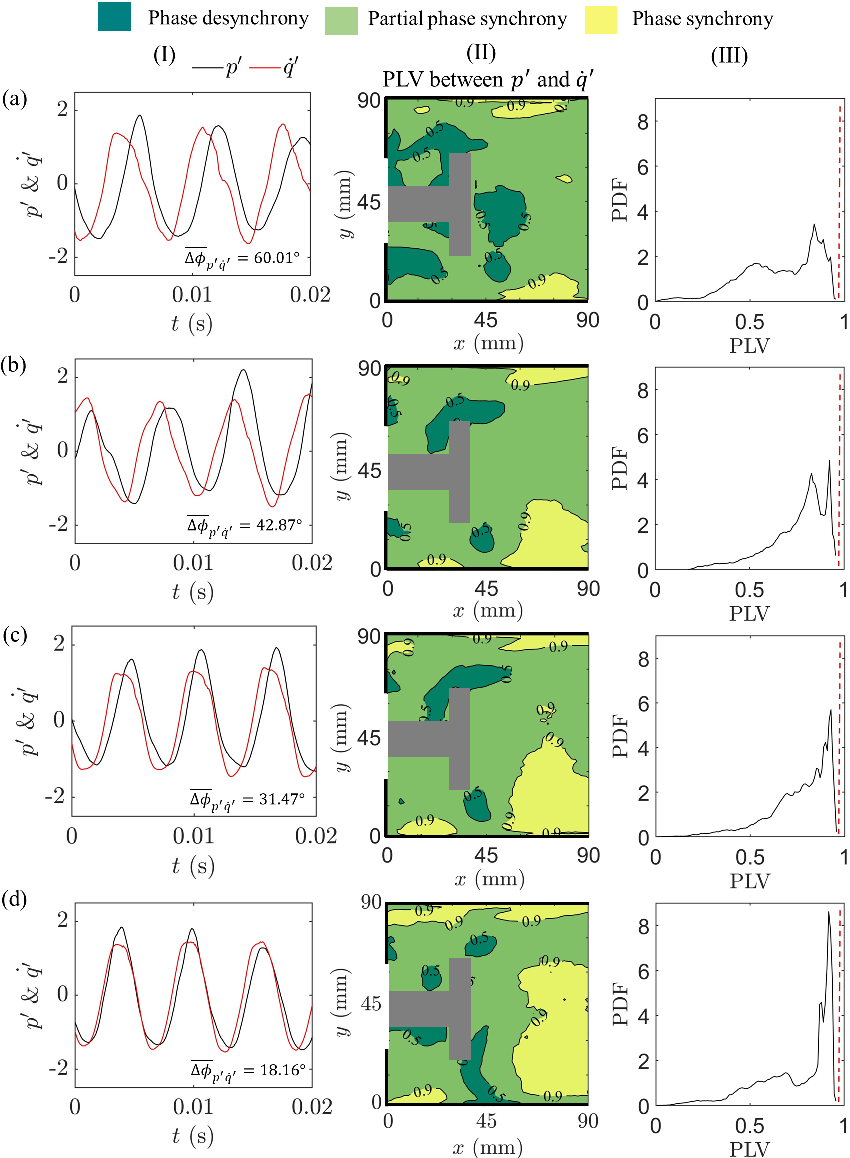}}
  \caption{(I) The normalized time series of $p^\prime$ and global  $\dot{q}^\prime$ oscillations. (II) The spatial contours of PLV between $p^\prime$ and local  $\dot{q}^\prime$ oscillations, categorizing regions of phase desynchrony (0 $\le$ PLV $\le$ 0.5), partial phase synchrony (0.5 $<$ PLV $<$ 0.9), and phase synchrony (0.9 $\le$ PLV $\le$ 1). (III) The probability density function (PDF) of spatially distributed PLV, where a red dashed line near 1 indicates the PLV observed for synchronized $p^\prime$ and global $\dot{q}^\prime$ oscillations shown in (I). The mean phase difference between $p^\prime$ and global $\dot{q}^\prime$ oscillations is indicated at the bottom right corner of each plot in (I). The plots in (a) to (d) are obtained for different states of thermoacoustic instability observed during the transition from phase synchronization to generalized synchronization, corresponding to the global equivalence ratio values of 0.68, 0.63, 0.6, and 0.54, respectively. We observe that the small pockets of synchronized $p^\prime$ and $\dot{q}^\prime$ oscillations (PLV $>$ 0.9) first originate near the combustor walls during the onset of phase synchronization (a). As the system transitions to generalized synchronization (b to d), these pockets of synchronized oscillations grow in size and spread to cover a substantial portion of the reaction field.}
\label{fig7}
\end{figure}


Next, we investigate the time-averaged phase interaction between $p^\prime$ and local $\dot{q}^\prime$ oscillations in the reaction field using phase locking value (PLV), focusing on the transition from the state of phase synchronization to generalized synchronization. As evident from Fig. \ref{fig4}a and Fig. \ref{fig7}-I, the global oscillations of $p^\prime$ and $\dot{q}^\prime$ exhibit perfect phase locking, resulting in the PLV remaining close to 1. However, intriguing patterns of phase locking emerge at the local scale (refer to Fig. \ref{fig7}-II). Given that the PLV ranges between 0 and 1, we classify it into three distinct ranges based on the overall phase-locking properties of $p^\prime$ and local $\dot{q}^\prime$ oscillations within the reaction zone. The identified regions of phase interaction are as follows: phase synchrony (0.9 $\le$ PLV $\le$ 1), partial phase synchrony (0.5 $<$ PLV $<$ 0.9), and phase desynchrony (0 $\le$ PLV $\le$ 0.5), which are indicated by distinct colors in Fig. \ref{fig7}-II.

We found that the regions of local phase desynchrony between $p^\prime$ and $\dot{q}^\prime$ oscillations nearly overlap with those of local frequency desynchronized oscillations identified in Fig. \ref{fig6}-II. These regions of local phase desynchrony persist during the transition from the state of phase synchronization to generalized synchronization of thermoacoustic instability. However, in the regions of frequency-synchronized $p^\prime$ and local $\dot{q}^\prime$ oscillations (seen in Fig. \ref{fig6}), we observe partial phase synchrony (0.5 $<$ PLV $<$ 0.9) and perfect phase synchrony (0.9 $\le$ PLV $\le$ 1) in the phase interaction of these oscillations (as shown in Fig. \ref{fig7}-II). During partial synchrony, intermittent phase locking of $p^\prime$ and local $\dot{q}^\prime$ oscillations is noticed over time at particular locations in the reaction field (plot not shown for brevity).

Our overall observations based on PLV calculation suggest that, during the state of phase synchronization of thermoacoustic instability, very small pockets of local phase-locked $p^\prime$ and $\dot{q}^\prime$ oscillations (i.e., PLV $\ge$ 0.9) are observed in the reaction field (Fig. \ref{fig7}a-II). These regions initially appear near the top and bottom walls of the combustor, eventually spreading along the combustor walls and growing towards the center of the reaction field (downstream of bluff-body), as the system transitions from phase synchronization to generalized synchronization of thermoacoustic instability (compare yellow regions in Figs. \ref{fig7}a-II to \ref{fig7}d-II). This widening of local regions of phase synchrony between $p^\prime$ and $\dot{q}^\prime$   oscillations in the reaction field is also evident from the increase in the height of the probability density function (PDF) of the spatially distributed PLV near unity (refer to Figs. \ref{fig7}a-III to \ref{fig7}d-III). Therefore, we discover that while the coupled behavior of $p^\prime$ and $\dot{q}^\prime$ oscillations maintains a global order (synchrony), i.e., PLV closer to 1, during the transition from phase synchronization to generalized synchronization of thermoacoustic instability (Fig. \ref{fig4}a), the underlying local phase coupling between these oscillations demonstrates complex dynamics, including the coexistence of synchrony, partial synchrony, and desynchrony in the reaction field (Fig. \ref{fig7}-II). 

 \begin{figure} 
  \centerline{\includegraphics[scale=0.85]{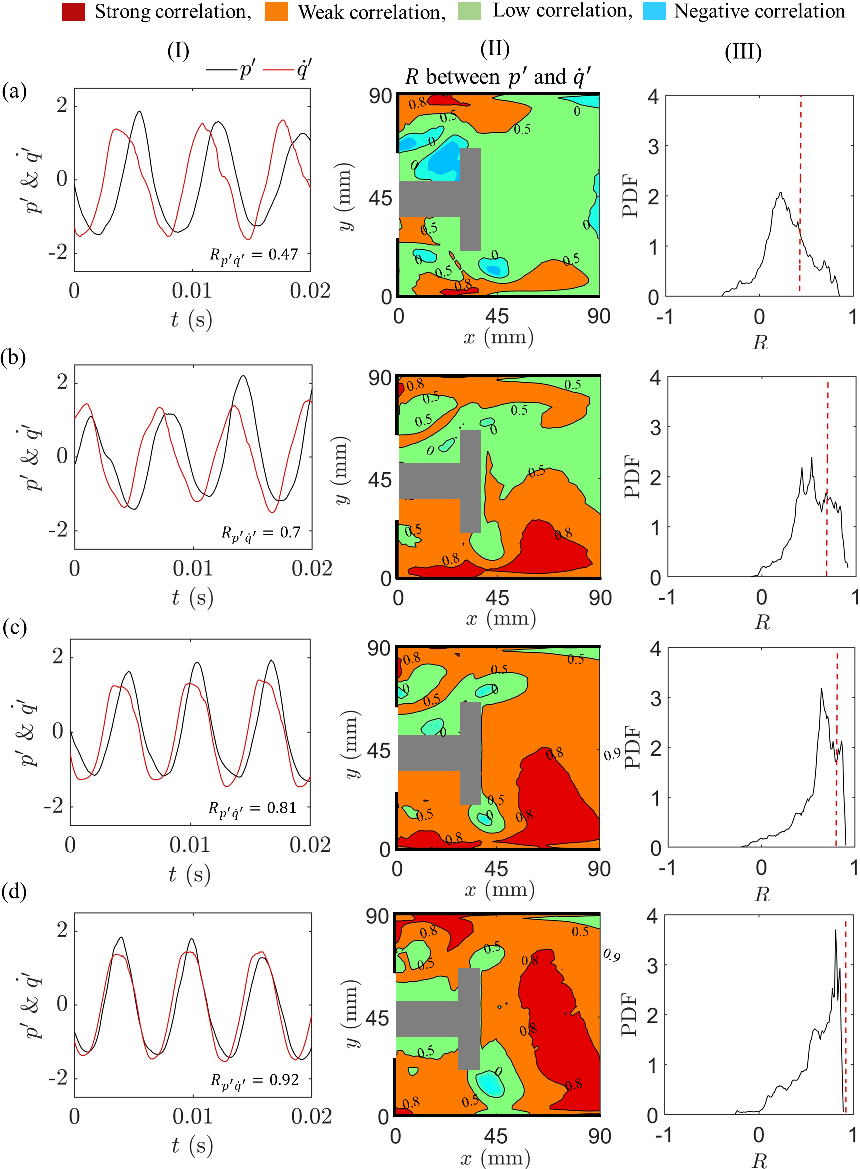}}
  \caption{(I) The normalized time series of $p^\prime$ and global  $\dot{q}^\prime$ oscillations. (II) The spatial contours of amplitude locking between $p^\prime$ and local $\dot{q}^\prime$ oscillations, highlighting regions of strong positive correlation ($0.8 \le R \le 1$), weak positive correlation ($0.5 < R<0.8$), low positive correlation ($0 \le R \le 0.5$), and negative correlation ($R<0$) in the reaction field. (III) The probability density function (PDF) of $R$ between $p^\prime$ and local  $\dot{q}^\prime$ oscillations in the reaction field, where the red dotted line indicates the value of $R$ between $p^\prime$ and global  $\dot{q}^\prime$ oscillations as shown at the bottom right corner of each plot in (I). The plots in (a) to (d) correspond to different states of thermoacoustic instability as discussed in Fig. \ref{fig7}. We observe that during phase synchronization, small local pockets of weakly and strongly correlated $p^\prime$ and $\dot{q}^\prime$ oscillations first originate near the combustor walls. These regions eventually expand and cover a significant portion of the reaction field during the occurrence of generalized synchronization.} 
\label{fig8}
\end{figure}

Having examined the local frequency and phase locking properties of the coupled $p^\prime$ and $\dot{q}^\prime$ oscillations within the reaction field, we now present their amplitude locking characteristics, determined through the Pearson correlation coefficient $R$ (see Eq. \ref{eq6}). Figure \ref{fig8}-II illustrates the spatial correlation regions of $R$ between $p^\prime$ and local $\dot{q}^\prime$ oscillations observed during various states of thermoacoustic instability, specifically focusing on the transition from phase synchronization to generalized synchronization. In Figs. \ref{fig4}b and \ref{fig8}-I, we notice that the value of $R$ between the $p^\prime$ and global $\dot{q}^\prime$ oscillations is 0.47 during the onset of phase synchronization, but it increases to 0.92 during the occurrence of generalized synchronization. We use these $R$ values as a reference for categorizing the spatial features of the amplitude correlation between $p^\prime$ and local $\dot{q}^\prime$ oscillations into three groups: strong correlation ($0.8 \le R \le 1$), weak correlation ($0.5 < R < 0.8$), and low correlation ($0 \le R \le 0.5$). Furthermore, we identify regions of positive correlation ($R > 0$) that included all low, weak, and strong correlation areas, as well as regions of negative correlation ($R<0$). These positive and negative regions corresponded to areas of acoustic driving and damping in the reaction field of the combustor, respectively \citep{lieuwen2005combustion}. All these regions are visually represented by different colors in Fig. \ref{fig8}-II.


During the onset of phase synchronization (Fig. \ref{fig8}a-II), we observe very small pockets of weakly and strongly correlated regions of $p^\prime$ and local $\dot{q}^\prime$ oscillations. These regions are concentrated near the top and bottom walls of the combustor. The majority of the reaction field is covered with low correlated oscillations. Consequently, the PDF of the spatially distributed values of $R$ (Figs \ref{fig8}a-III) shows a peak around 0.25, which is lower than the value of $R=0.47$ observed between $p^\prime$ and global $\dot{q}^\prime$ oscillations (indicated by a red dotted line in Fig. \ref{fig8}a-III) during this state. Negatively correlated regions of $p^\prime$ and local $\dot{q}^\prime$ oscillations are also observed in the shear layers that are stabilized at the dump plane and on the shaft and the tip of the bluff-body. As the system transitions to the state of generalized synchronization (compare Figs. \ref{fig8}a-II to \ref{fig8}d-II), we observe a significant growth in the pockets of weakly and strongly correlated $p^\prime$ and local $\dot{q}^\prime$ oscillations in the reaction field. These pockets progressively spread toward the center of the combustor and eventually cover the majority of the reaction field. Consequently, the peak of the PDF of $R$ moves towards 1 during the transition from phase synchronization to generalized synchronization (compare Figs. \ref{fig8}a-III to \ref{fig8}d-III). Furthermore, we note that the regions of negatively correlated $p^\prime$ and local $\dot{q}^\prime$ oscillations persist in the reaction field throughout the transition from phase synchronization to generalized synchronization.  
 
As we previously discussed in equations (\ref{eq6}) and (\ref{eq7}), the Pearson correlation coefficient $R$ is the same same as the normalized Rayleigh index $RI$. Using the Rayleigh index to describe our results, we can infer that during the state of generalized synchronization, the spatial acoustic driving caused by local heat release rate fluctuations is more intense than during phase synchronization. This stronger driving leads to a greater coupling between $p^\prime$ and local $\dot{q}^\prime$ oscillations, which is reflected in the wider distribution of phase locking values (PLV) shown in Fig. \ref{fig7}-II. As a result of this increased coupling, the amplitude of pressure oscillations (as seen in Fig. \ref{fig2}e) is significantly higher during the state of generalized synchronization than during phase synchronization of thermoacoustic instability. 

\subsection{Spatial patterns of flow-acoustic interaction and critical regions}

\begin{figure}
  \centerline{\includegraphics[scale=1]{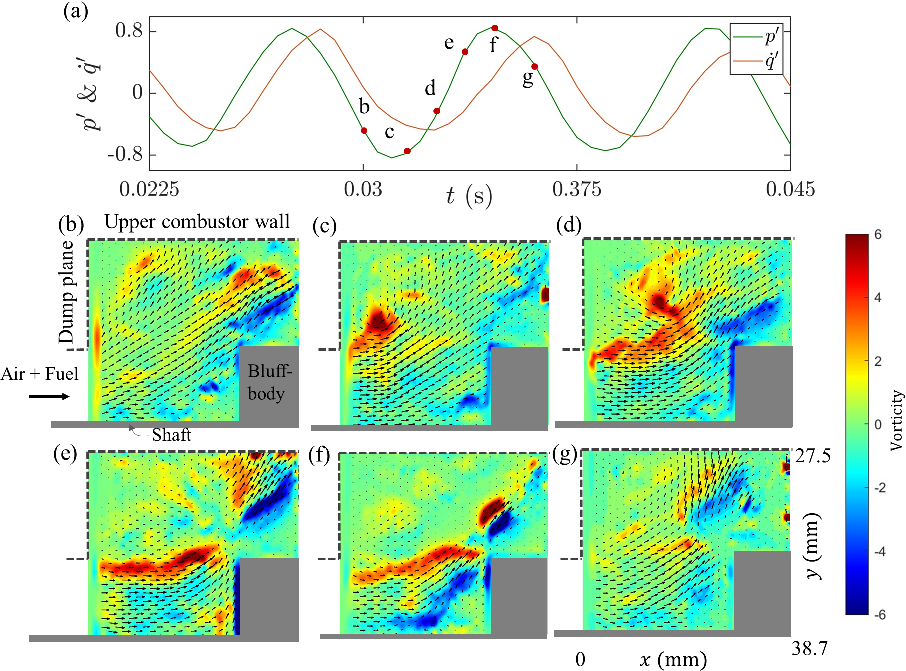}}
  \caption{(a) The normalized time series of acoustic pressure ($p^\prime$) and global heat release rate ($\dot{q}^\prime$) signals observed during the state of phase synchronization. Plots (b) to (g) show the instantaneous vorticity superimposed with the velocity field observed over a cycle of $p^\prime$ signal. The black dotted line indicates the combustor boundary, while the solid grey region denotes the bluff-body and the shaft. During the phase synchronization state, a small-sized vortex emerged at the corner of the dump plane, with its vorticity primarily concentrated along the outer shear layer.  }
\label{fig9}
\end{figure}

In the preceding discussion, we examined the differences in the spatiotemporal coupling between the heat release rate fluctuations in the flame and the acoustic pressure fluctuations in the combustor for the states of phase synchronization and generalized synchronization. Here, we proceed to investigate the characteristic hydrodynamic fluctuations underlying these states by analyzing the flow field acquired using Particle Image Velocimetry (PIV) measurements \citep{raffel1998particle}. Figures \ref{fig9}b-\ref{fig9}g display the instantaneous velocity field superimposed on the vorticity field at various instances during one oscillation cycle of the $p^\prime$ signal (Fig. \ref{fig9}a) for the state of phase synchronization. The X-Y axis in each plot (Figs. \ref{fig9}b-g) corresponds to the physical dimensions of the upper half of the combustor located between the dump plane and the bluff-body, with the flow direction from left to right. Unni \textit{et al.} \cite{unni2017flame}, using planar Mie scattering imaging of seeded fluid particles in the same combustor, previously identified two shear layers in the system, referred to as the outer and inner shear layers, stabilized at the tip of the dump plane and on the shaft of the bluff-body, respectively. The blue and red contours in Figs. \ref{fig9}b-g indicate the vorticity fields along these shear layers, corresponding to positive vorticity (counter-clockwise rotating vortex, +$\omega$) and negative vorticity (clockwise rotating vortex, -$\omega$), respectively. In addition, the black arrows denote the direction and magnitude of the local velocity field within the combustor.

Numerous previous studies \citep{rogers1956mechanism,poinsot1987vortex,smith1985combustion,schadow1992combustion,chakravarthy2017dynamics,oconnor2022understanding} have shown that during the state of thermoacoustic instability, strong vortex-acoustic synchrony is established in the combustor, with large-scale vortical structures emerging periodically at the same frequency as the acoustic fluctuations. During the state of phase synchronization (Fig. \ref{fig9}), we observe that upon entry of the fresh reactant mixture into the combustor, the sudden expansion of the flow passage leads to the formation of a small-sized vortex with positive vorticity at the tip of the dump plane. The reactant mixture splits, with a portion carried by the vortex, some moving along the shear layer, and the remainder passing along the shaft of the bluff body. These reactants subsequently mix with hot products already present in the reaction zone from the previous oscillation cycle. The presence of the vortex leads to a roll-up of fluid particles along the outer shear layer. Over time, the vortex grows in size as it convects downstream toward the bluff-body and the wall of the combustor (Fig. \ref{fig9}d).  

We notice that the reactant mixture moving along the shaft accelerates towards the bluff body during the vortex roll-up process (Fig. \ref{fig9}d) and subsequently impinges upon the bluff body before reflecting backward (Figs. \ref{fig9}e,f). This vortex impingement enhances the mixing between the hot products and cold reactants, leading to a localized region of high heat release rate upstream of the bluff body. At the same time, the vortex disintegration process at the top wall of the combustor (Fig. \ref{fig9}e,f) leads to the formation of many small-scale structures. A portion of these structures moves along the wall towards the dump plane, while the remaining structures entrain into the flow moving along the shear layer (Fig. \ref{fig9}g,b). During the vortex breakdown, an intense mixing of cold reactants carried by the vortex and hot products left over from the previous cycle creates a localized region of high heat release rate near the top wall. It is important to note that the two events of heat release discussed here, namely behind the bluff body and near the top wall of the combustor, occur simultaneously. Previously, Seshadri \textit{et al.} \cite{seshadri2016reduced} considered such localized heat release due to the impingement of a vortex on the bluff-body as a ‘kick’. This kick generates large amplitude acoustic waves that propagate upstream, affecting the vortex formation process at the dump plane \citep{pawar2017thermoacoustic}. As pressure amplitude drops, the strength of positive vorticity reduces due to dissipation (Fig. \ref{fig9}g,b). Finally, the remaining reactants get consumed, and the combustion products are pushed over the bluff body along the shear layer, paving the way for the next oscillation cycle. The flow passing over the bluff body induces negative vorticity on the top of the bluff body (Figs. \ref{fig9}d-f). 

\begin{figure}
  \centerline{\includegraphics[scale=1]{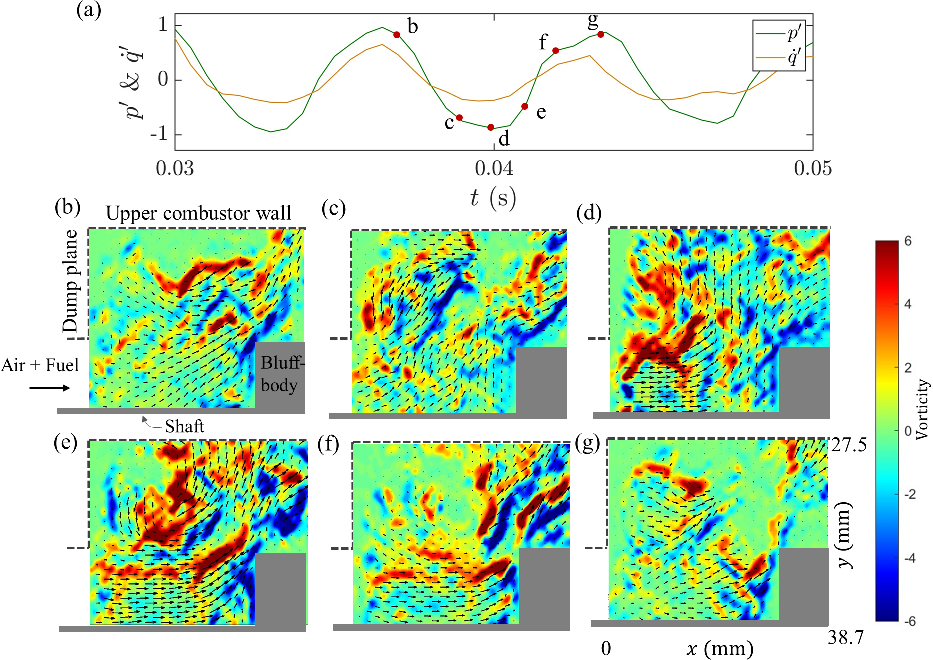}}
  \caption{(a) The normalized time series of acoustic pressure ($p^\prime$) and global heat release rate ($\dot{q}^\prime$) observed during the state of generalized synchronization. Plots (b) to (g) show the instantaneous vorticity superimposed with the velocity field observed over a cycle of $p^\prime$ signal. In the generalized synchronization state, the size of the vortex with positive vorticity emerging from the dump plane is larger, and the vorticity is distributed almost throughout the flow field. }
\label{fig10}
\end{figure}

Figure \ref{fig10} depicts the instantaneous spatiotemporal flow field corresponding to a single cycle of $p^\prime$ fluctuations during the state of generalized synchronization. The sequence of processes of vortex formation (Fig. \ref{fig10}c), convection (Fig. \ref{fig10}d,e), and impingement (Fig. \ref{fig10}f) resulting in localized heat release rate fluctuations in the combustor is similar to that discussed for the state of phase synchronization (Fig. \ref{fig9}). However, some key differences are noticeable between these states. Specifically, the vortex size formed at the dump plane during each oscillation cycle is larger for the state of generalized synchronization than for the state of phase synchronization (compare Fig. \ref{fig9}d and \ref{fig10}d). The larger vortex size and higher flow velocity in the state of generalized synchronization lead to vortex impingement on the combustor wall, closer to the dump plane, when compared to the state of phase synchronization where it occurs near the bluff-body (refer to Fig. \ref{fig9}e and \ref{fig10}e). Additionally, the magnitude of the local positive vorticity observed for the state of generalized synchronization is higher and spread over the entire reaction zone (Fig. \ref{fig10}c-f), whereas for the state of phase synchronization, it is lower and predominantly observed along the outer shear layer (Fig. \ref{fig9}c-f).

\begin{figure}
  \centerline{\includegraphics[scale=0.9]{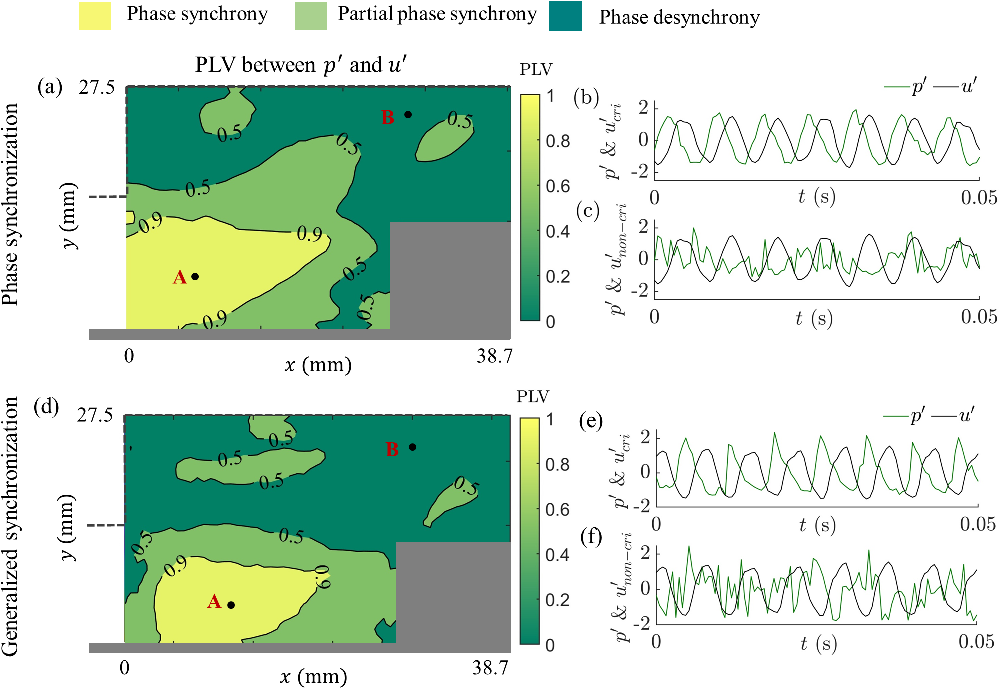}}
\caption{(a, d) The spatial contours of PLV between $p^\prime$ and local $u^\prime$ oscillations, highlighting regions of phase synchrony (0.9 $\le$ PLV $\le$ 1), partial phase synchrony (0.5 $<$ PLV $<$ 0.9), and desynchrony (0 $\le$ PLV $\le$ 0.5), for the state of phase synchronization and generalized synchronization, respectively. Plots in (b, e) and (c, f) represent the time series of $p^\prime$ fluctuations and $u^\prime$ fluctuations in the critical region ($u_{cri}$) and the non-critical region ($u_{non-cri}$) corresponding to spatial points A and B, respectively, as indicated both in (a) and (d). The critical region is located at the combustor entrance over the shaft of the bluff-body, while the non-critical region is spread along the combustor wall and the top of the bluff-body. $u^\prime$ oscillations are periodic in the critical region, while they are aperiodic in the non-critical region. }
\label{fig11}
\end{figure}
  
Next, we quantify the local flow-acoustic coupling in the reaction field for the state of phase synchronization (Fig. \ref{fig11}a) and generalized synchronization (Fig. \ref{fig11}d) of thermoacoustic instability. To accomplish this, we use PLV to examine the local synchronization of the acoustic pressure $p^\prime$ (measured near the dump plane) and the local streamwise velocity $u^\prime$ fluctuations in the reaction field (obtained through PIV measurements), similar to what we did to study the coupling between $p^\prime$ and local  $\dot{q}^\prime$ fluctuations in Fig. \ref{fig7}. Depending on the synchronization characteristics of $p^\prime$ and local $u^\prime$ fluctuations, we classify the spatial regions of PLV into phase synchrony (0.9 $\le$ PLV $\le$ 1), partial phase synchrony (0.5 $<$ PLV $<$ 0.9), and desynchrony (0 $\le$ PLV $\le$ 0.5). Our observations reveal that for both states of synchronization, the region of phase-locked $p^\prime$ and $u^\prime$ oscillations (i.e., PLV close to 1) is primarily concentrated near the shaft of the bluff-body where the flow enters the combustor (refer to Fig. \ref{fig11}a and \ref{fig11}d). This region is surrounded by partially synchronized $p^\prime$ and local $u^\prime$ oscillations (i.e., 0.5 $<$ PLV $<$ 0.9). The regions of desynchronized $p^\prime$ and local $u^\prime$ oscillations are observed along the combustor wall and also over the top of the bluff-body. Thus, we notice that the local phase interaction of $p^\prime$ and local $u^\prime$ oscillations exhibits a coexistence of phase synchrony and desynchrony in the reaction field during thermoacoustic instability, as also witnessed in the local phase interaction of $p^\prime$ and local  $\dot{q}^\prime$ fluctuations in Fig. \ref{fig7}.


We refer to the regions of the reaction zone where $p^\prime$ and $u^\prime$ oscillations are synchronized (i.e., 0.9 $\le$ PLV $\le$ 1) as the \textit{critical} regions and those where $p^\prime$ and $u^\prime$ oscillations are desynchronized (i.e., 0 $\le$ PLV $\le$ 0.5) as the \textit{non-critical} regions. A critical region represents the highly sensitive zone in the reaction field of a turbulent combustor that is likely responsible for the generation and sustenance of thermoacoustic instabilities \citep{unni2018emergence,krishnan2019mitigation}. Within the critical region, we observe that $u^\prime$ oscillations are periodic (see Figs. \ref{fig11}b and \ref{fig11}e), while they are aperiodic in the non-critical region (see Figs. \ref{fig11}c and \ref{fig11}f). Therefore, to suppress thermoacoustic instabilities, we believe that breaking the strong synchrony between the flow and the acoustic subsystems in the reaction field is essential. This can be achieved by disrupting such patterns of the critical region through external perturbations, leading to an effective methodology for locating actuators on the combustor. 

Numerous previous studies have highlighted the significance of different regions in a reactive flow field in governing the spatiotemporal dynamics of the turbulent combustor during thermoacoustic instability. Uhm \textit{et al.} \cite{uhm2005low} identified the region of local heat release rate maxima as the optimal region, while Ghoniem \& co-workers \citep{ghoniem2005stability,altay2010mitigation} and Murayama \textit{et al.} \cite{murayama2019attenuation} considered the flame anchoring location as the optimal region to control thermoacoustic instabilities. Tachibana \textit{et al.} \cite{tachibana2007active} used a distribution of the Rayleigh index to optimize for the choice of the secondary fuel injector. Recently, Unni \textit{et al.} \cite{unni2018emergence} used a complex network approach to identify critical regions in the combustor based on the correlation between velocity fields, while Krishnan \textit{et al.} \cite{krishnan2019mitigation} showed that targeting regions with large network measure values obtained from the flow field is the most effective and efficient way to control thermoacoustic instability. Similarly, Tandon \& Sujith \cite{tandon2023multilayer} used the multilayer network approach to identify critical regions (hubs) through the interaction between thermoacoustic power and vorticity field.  On the other hand, Roy \textit{et al.} \cite{roy2021critical} found that the Hurst exponent calculation for turbulent velocity fluctuations is a more robust measure for identifying critical regions for suppressing thermoacoustic instability than maxima in the Rayleigh index and averaged heat release rate field in the flame. We would like to highlight that the aforementioned studies primarily concentrated on identifying critical regions through data-based analysis of pressure, heat release rate, and flow velocity fluctuations. In contrast, our study takes a different approach, focusing on the detection of critical regions within the reaction field by examining the local synchrony between pressure and velocity fluctuations. This particular aspect of interaction is crucial in understanding and predicting the onset of thermoacoustic instability. Additionally, we note that none of the previous studies discussed the suppression of different forms of thermoacoustic instability, such as phase synchronization and generalized synchronization, through critical region identification, which we examine in the subsequent discussion. 

\begin{figure}
  \centerline{\includegraphics[scale=1.2]{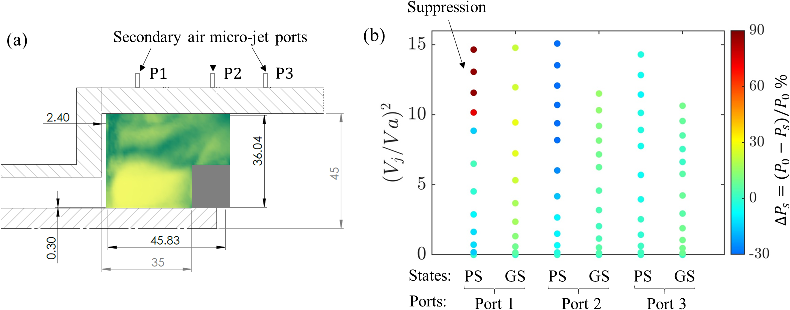}}
  \caption{(a) The schematic of the experimental arrangement illustrating the injection of secondary air through ports P1, P2, and P3 in the reaction field to enable smart control of thermoacoustic instability. (b) A scatter plot displaying the relative change in the root-mean-square value of $p^\prime$ oscillations, $\Delta P$, as a function of the total momentum flux ratio, $(V_j/V_a)^2$, when a micro-jet is injected through a particular port during the state of phase synchronization (PS) and generalized synchronization (GS). We observe complete suppression only for the phase synchronization state of thermoacoustic instability when micro-jet injection targets the critical region in the combustor.}
\label{fig12}
\end{figure}

In Fig. \ref{fig12}a, we illustrate a smart control methodology using secondary air injection to suppress thermoacoustic instability in a turbulent combustor. A similar methodology has been used previously by a few researchers \citep{ghoniem2005stability,altay2010mitigation,murayama2019attenuation,zhou2020optimal,krishnan2019mitigation,deshmukh2017experiments} to disturb the flow and flame dynamics locally in the combustor and thus control thermoacoustic instabilities. To achieve  secondary air injection, we strategically install small diameter ports on both the top and bottom walls of the combustor at specific locations, such that the air micro-jets are directed towards the critical region through port P1, on top of the bluff-body through port P2, and downstream of the bluff-body through port P3 (see Fig. \ref{fig12}a). The flow rate of air supplied through these ports is lesser than 10 times the flow rate of the primary air entering the combustor through the burner section. The small size of the ports creates secondary air injection at high-speed that, in turn, generates strong local disturbances in the flow field. When the air micro-jet of a particular flow velocity is allowed to impinge on the appropriate location in the combustor, it breaks the local flow-flame-acoustic coupling and mitigates thermoacoustic instability \citep{krishnan2019mitigation,roy2021critical}.    

Figure \ref{fig12}b shows the percentage change in the amplitude of $p^\prime$ oscillations, denoted as $\Delta P_s$, in relation to the change in total momentum flux ratio, $(V_j/V_a)^2$, of two jets within the combustor. Here, $\Delta P_s$ represents the normalized change in the RMS value of $p^\prime$ oscillations caused by micro-jet injection, calculated as: $\Delta P_s=(P_0-P_s)/P_0$, where $P_0$ corresponds to the RMS value of $p^\prime$ oscillations without air injection. Furthermore, $V_j$ refers to the velocity of the micro-jet through each port, while $V_a$ represents the velocity of the reactant mixture entering the combustor. In each experiment, we first establish limit cycle oscillations of amplitude ($P_0$) corresponding to a particular state of thermoacoustic instability, and then progressively increase the velocity of micro-jets ($V_j$) by varying the air flow rates. We investigate the impact of micro-jet air injection through ports P1, P2, and P3 on the suppression of $p^\prime$ oscillations in both the phase synchronization (PS) and generalized synchronization (GS) states.  

When we inject air micro-jets through port P1, specifically targeting the critical region shown in Figs. \ref{fig11}a and \ref{fig11}d, we observe the suppression of thermoacoustic instability during the state of phase synchronization (as illustrated in the first case of Fig. \ref{fig12}b). The amplitude of $p^\prime$ oscillations is reduced to approximately 86\% of its original value, reaching a level comparable to the pressure amplitude observed for combustion noise, when $(V_j/V_a)^2$ exceeds a particular value. In contrast, during the state of generalized synchronization, we only observe a small decrease in the amplitude of $p^\prime$ oscillations, specifically less than 30\%, due to injecting micro-jets through port P1. However, when the micro-jet air injection is directed through ports P2 and P3, which are located away from the critical region in the reaction field, we do not observe the suppression of $p^\prime$ oscillations for any of the synchronization states of thermoacoustic instability. Instead, we observe the amplification of these oscillations, particularly during the state of phase synchronization, as indicated by negative values of $\Delta P_s$ in Fig. \ref{fig12}b. 

Our findings demonstrate that the suppression of thermoacoustic instability, particularly during the state of phase synchronization, is relatively easy and consistently observed in almost all experimental runs. The weak interaction between the flame-flow-acoustic subsystems during this state can be easily disrupted by eliminating the critical region within the reaction field through micro-jet injection, thereby achieving effective suppression of thermoacoustic instability. In particular, by targeting the critical regions with micro-jets, we prevent the development of collective behavior between the small and large-scale vortices along the outer shear layer, a behavior typically observed during thermoacoustic instability \citep{george2018pattern}. Consequently, the formation of large-scale structures is hindered, leading to the suppression of thermoacoustic instability. In contrast, despite targeting the critical regions, the micro-jets are unable to disrupt a larger vortex and the stronger flame-flow-acoustic interaction that develops during the state of generalized synchronization. As a result, the suppression of thermoacoustic instability is less effective during this state. 


\section{Conclusions}

To conclude, in the present study we examine the characteristics of both global and local coupling between the flame, flow, and acoustic fields of a bluff-body stabilized turbulent combustor during phase synchronization and generalized synchronization states of thermoacoustic instability. To that end, we use the framework of synchronization theory and analyze the coupled behavior of three self-sustained oscillators of the combustor: the acoustic field of the combustor, the heat release rate fluctuations in the flame, and the velocity fluctuations in the turbulent flow. Various measures of synchrony are employed to quantify the coupling in frequency (spectral coherence), phase (phase locking value, PLV), and amplitude (Pearson correlation coefficient) of these oscillators. 

We observe a weak form of limit cycle oscillations during phase synchronization and a strong form of limit cycle oscillations during generalized synchronization of acoustic pressure and global heat release oscillations. The measures of synchronization exhibit a smooth variation as the system transitions from a state of phase synchronization to generalized synchronization. During this transition, an increase in the natural frequency of pressure and global heat release oscillations is accompanied by a decrease in the mean phase difference between them towards zero and a corresponding increase in the correlation coefficient towards unity. Upon the onset of generalized synchronization, the variation in these measures becomes very small.

We discovered that the spatiotemporal interaction between the flow, flame, and acoustic subsystems in a spatially extended turbulent combustor is complex at the local scale despite the ordered period-1 thermoacoustic oscillations being observed at the global scale. Various patterns of spatial interaction between these subsystems are identified by studying the coupling of frequency, phase, and amplitude of local oscillations in the reaction field. We found that regions of frequency synchronized and desynchronized acoustic pressure and local heat release rate oscillations coexist in the reaction field during both phase synchronization and generalized synchronization states of thermoacoustic instability. Moreover, various phase locking patterns, such as perfect phase synchrony and partial phase synchrony, are witnessed in regions of frequency synchronization of these oscillations. We observed that the formation of small pockets of phase synchrony between the acoustic pressure and local heat release rate oscillations is sufficient for the onset of phase synchronization at the global scale. These pockets of synchrony then gradually grow and spread in the reaction field during the transition from phase synchronization to generalized synchronization. The spatial field of amplitude correlation between these oscillations exhibits a few regions of negatively correlated oscillations among largely populated regions of positive correlation. These positively correlated oscillations are further classified into weak correlation and strong correlation, both of which grow in size from small pockets to large islands during the transition from phase synchronization to generalized synchronization. 

Finally, we studied the local synchronization characteristics of acoustic pressure and stream-wise velocity fluctuations in the flow using the phase locking value. We observed the coexistence of distinct regions of perfect phase synchrony, partial phase synchrony, and desynchrony of these oscillations in the reaction field for both phase synchronization and generalized synchronization states of thermoacoustic instability, similar to the local synchronization patterns observed between acoustic pressure and heat release rate fluctuations. Interestingly, we observed that synchronized pressure and velocity oscillations occur at the combustor entrance along the shaft of the bluff-body, in contrast to the synchronized acoustic pressure and local heat release rate fluctuations that are predominantly located near the combustor wall and far downstream of the bluff-body. We referred to the regions of synchronized acoustic pressure and local velocity fluctuations as critical regions within the combustor flow field. The destruction of such regions through a smart control strategy, achieved by injecting secondary air micro-jets, effectively suppressed phase synchronized thermoacoustic oscillations in the system.  

Thus, our spatiotemporal analysis of a turbulent thermoacoustic system indicates that an intrinsic balance between order, partial order, and disorder emerges in the coupled flow, flame, and acoustic subsystems at the local scale to maintain an ordered behavior at the global scale during thermoacoustic instability. This interplay between subsystems during thermoacoustic instability suggests that a turbulent combustion system is a complex system, and can be deeply understood through a complex system approach \citep{sujith2021thermoacoustic}. Complex systems are characterized by many components, where a local nonlinear interaction between these components leads to the emergence of global ordered patterns \citep{anderson1972more, kauffman1993origins, bar1997dynamics, parisi1999complex,thurner2018introduction}. 
 Examples of other complex systems include the human body, ecosystems, the stock market, traffic jams, web services, and self-propelled particles \citep{bar1997dynamics,auyang1998foundations,haken2006information,sneyd2001self,sumpter2010collective}. The spatially extended turbulent combustion systems exhibit many common features of a typical complex system, such as the nonlinear delayed interaction between subsystems, the emergence of order from disorder, self-organization of small-scale vortices into a large-scale vortical structure, and collective behavior and synchronization \citep{sujith2021thermoacoustic}. Therefore, understanding the spatiotemporal patterns in the interaction between the flame, flow, and acoustic fluctuations at both local and global scales through different approaches from complex system theory can aid in developing efficient control strategies to mitigate thermoacoustic instabilities.

\section*{Acknowledgement}
R.I.S. gratefully acknowledges funding from J. C. Bose Fellowship offered by the Department of Science and Technology, India (No.JCB/2018/000034/SSC).


\bibliography{apssamp}

\end{document}